\documentclass[12pt]{iopart}
\usepackage{graphicx}
\usepackage{iopams}
\begin{document}
\title[The maximal acceleration and ERD]{The maximal acceleration, Extended Relativistic Dynamics and Doppler type shift for an accelerated source}

\author{ Y. Friedman }
\address{Jerusalem College of
Technology\\P.O.B. 16031 Jerusalem 91160, Israel\\email: friedman@jct.ac.il}

\begin{abstract}
Based on the generalized principle of relativity and the ensuing symmetry, we have shown that there are only two possible types of transformations between uniformly accelerated systems. The first allowable type of transformation  holds if and only if the Clock Hypothesis is true. If the Clock Hypothesis is not true, the transformation is of Lorentz-type and implies the existence of a universal maximal acceleration $a_m$.

We present an extension of relativistic dynamics for which all admissible solutions will have have a speed bounded by the speed of light $c$ and the acceleration bounded by $a_m$. An additional Doppler type shift for an accelerated source is predicted. The formulas for such shift are the same as for the usual Doppler shift with $v/c$ replaced by $a/a_m$.

 The W. K\"{u}ndig experiment of measurement of the transverse Doppler shift in an accelerated system was also exposed to a longtitudal shift due to the acceleration. This experiment, as reanalyzed  by Kholmetskii et al, shows that the Clock Hypothesis is not valid. Based on the results of this experiment, we predict that the value of the maximal acceleration $a_m$ is of the order $10^{19}m/s^2$. Moreover, our analysis provides a way to measure experimentally the maximal acceleration with existing technology.


\end{abstract}

 \maketitle

\section{Transformation between uniformly accelerated systems}

The Lorentz transformations of special relativity can be derived from the principle of special relativity alone, \textit{without} assuming the constancy of the speed of light (see Friedman(2004), ch. 1). This approach was applied to the transformations between two \textit{uniformly accelerated} systems in Friedman and Gofman (2004) and (2010). The resulting description follows from the generalized principle of relativity and the ensuing symmetry.

We replace the space-time description of events by the proper velocity-time description. The proper velocity $\mathbf{u}$ of an object is the derivative of the object's displacement
with respect to the proper time. In other words, $\mathbf{u}=\gamma \mathbf{v}$, where $\mathbf{v}$ is the object's velocity and $\gamma({v})=1/\sqrt{1-\mathbf{v}^2/c^2}$. A system $K'$ is called \textit{uniformly accelerated} with respect to an inertial system $K$ if it moves parallel to it with constant acceleration $\mathbf{a}=d\mathbf{u}/dt$
(see  Moller (1972)and Franklin (2010)). In the proper velocity-time description of events, the transformation between uniformly accelerated systems become linear. We have shown, Friedman and Gofman (2010), that there are only two types of transformations between uniformly accelerated systems.

The validity of the \textit{Clock Hypothesis} is crucial to determining which one of the two types of transformations is obtained. The Clock Hypothesis in Einstein (1911) states that the rate of an accelerated clock is equal to that of a comoving unaccelerated clock. If the Clock Hypothesis is valid, then the transformations between uniformly accelerated systems are Galilean. If not, these transformations are of Lorentz-type, and, moreover, there exists a universal maximal acceleration, which we denote by $a_m$.

If the Clock Hypothesis is false, then the transformation between uniformly accelerated systems depends only on the relative acceleration $\mathbf{a}$ between the systems. If the acceleration $\mathbf{a}$ is chosen in the direction of the $x$-axis, the proper velocity-time transformations are
\begin{equation}\label{coordtranspfinal2}
  \begin{array}{cl}
    t &=\tilde{\gamma}(t'+ \frac{au'_x}{a_m^2}) \\
    u_x & =\tilde{\gamma}(at'+u'_x) \\
    u_y & =u'_y \\
    u_z & =u'_z,
  \end{array}
\end{equation}
where
\begin{equation}\label{gama_accel}
 \tilde{\gamma}=1/\sqrt{1-\frac{a^2}{a_m^2}}
\end{equation}
is the time dilation due to the acceleration.
The form $ (a_mt)^2-|\mathbf{u}|^2$ is an invariant for uniformly accelerated systems. This implies that the acceleration $a$ of any massive object is limited by the maximal one $a<a_m$. Moreover, the proper velocity time trajectory of such objects is inside the "light cone" $ |\mathbf{u}|<a_m t$ in the proper velocity time continuum. As in special relativity, the radiation generated by non-massive particles, will propagate on the boundary of the cone and thus be described by $f(\omega t- \tilde{k} u)$, with $|\tilde{k}|/\omega=1/a_m$.

\section{Extended Relativistic Dynamics}

The main feature of Special Relativity is that the set of all relativistically
allowed velocities is a ball $D_v \subset \mathbb{R}^3$ of radius $c$, the speed
of light. The Relativistic Dynamics (RD) describe motions which preserve this limitation. In Special Relativity the magnitude of the acceleration is unlimited, and the Clock Hypothesis holds. But, if the Clock Hypothesis is not true, the allowed accelerations belong to a ball $D_a \subset \mathbb{R}^3$ of radius $a_m$.
We extend RD to an \textit{Extended  Relativistic Dynamics }(ERD), a theory in which the speed of any moving object
is limited  by $c$ and the magnitude of its acceleration is limited by $a_m$.

 Newton's classical dynamics law $\mathbf{F}=m\mathbf{a}=m\frac{d\mathbf{v}}{dt}$ can be rewritten as
 \begin{equation}\label{class dynam}
   \left\{
    \begin{array}{ll}
      \frac{d\mathbf{x}}{dt}=\frac{\mathbf{p}}{m} \\
      \frac{d\mathbf{p}}{dt}=\mathbf{F}
    \end{array}\right.
  \end{equation}
where $\mathbf{p}=m \mathbf{v}$ is the momentum of the moving object.

The Relativistic Dynamics equation (Rindler (2004)) is $\mathbf{F}= m_0 \frac{d\mathbf{u}}{dt}$
where $\mathbf{u}=\gamma({v})\mathbf{v}$
is the proper velocity of the object and $m_0$ is the rest-mass of the object. In Friedman (2007)
it was shown that this dynamics equation can be derived from the boundness of the speed of a
moving object by $c$. Introducing the relativistic momentum $\mathbf{p}=m_0 \mathbf{u}=m_0\gamma({v})\mathbf{v}$ the RD equation can be rewritten as
\begin{equation}\label{RDyn}
 \left\{
    \begin{array}{ll}
     \gamma \left(\frac{d\mathbf{x}}{dt}\right)\frac{d\mathbf{x}}{dt}=\frac{\mathbf{p}}{m_0} \\
      \frac{d\mathbf{p}}{dt}=\mathbf{F}
    \end{array}
  \right.
\end{equation}

Note that RD equation brake the Born's Reciprocity, which state `` The laws of Nature are symmetrical with regard to space and momentum". In the above system, the derivative of the space in the first equation is by the proper time (including the time dilation due to the velocity of the moving object), while the derivative of the momentum in the second equation is by the lab time. In ERD we introduce in the second equation  the time
 dilation by a factor due to the acceleration
 of the object $\tilde{\gamma}({a})$, defined by (\ref{gama_accel}), with  $\mathbf{a}=d\mathbf{u}/dt$.

 Thus, the ERD equation  become
 \begin{equation}\label{ERD original}
   \left\{
    \begin{array}{ll}
     \gamma \left(\frac{d\mathbf{x}}{dt}\right)\frac{d\mathbf{x}}{dt}=\frac{\mathbf{p}}{m_0} \\
     \tilde{\gamma}\left(\frac{d\mathbf{u}}{dt}\right) \frac{d\mathbf{p}}{dt}=\mathbf{F}
    \end{array}
  \right.
 \end{equation}
 This equation can be derived also from boundness of the acceleration, as the RD equation is derived
  in Friedman (2007) from the boundness of the velocity.

\section{Hamilton type equations for ERD}

To obtain a Hamilton type equations for ERD, we define now for each, $\alpha\in\mathbb{R}$, a function $f_c:\mathbb{R}^3\rightarrow\mathbb{R}^3$ and its inverse by
\begin{equation}\label{f def}
    f_\alpha(\mathbf{x}) =\frac{\mathbf{x}}{\sqrt{1-|x|^2/\alpha^2}},
\;\;\;f_\alpha^{-1}(\mathbf{x}) =\frac{\mathbf{x}}{\sqrt{1+|x|^2/\alpha^2}}\,.
\end{equation}
With this notation, denoting $m=m_0$, we can rewrite (\ref{ERD original}) as
\[f_c\left(\frac{d\mathbf{x}}{dt}\right)=\mathbf{u},\;\;f_{a_m}\left(\frac{d\mathbf{u}}{dt}\right)=\frac{\mathbf{F}}{m}\]
or
\[\frac{d\mathbf{x}}{dt}=f_c^{-1}(\mathbf{u}),\;\;\frac{d\mathbf{u}}{dt}=f_{a_m}^{-1}
\left(\frac{\mathbf{F}}{m}\right)\,.\]

This turns the ERD equation into a non-linear first order system
of differential equations:
\begin{equation}\label{EST dym system}
    \left\{
      \begin{array}{l}
       \frac{d\mathbf{x}}{dt} =\frac{\mathbf{u}}{\sqrt{1+|\mathbf{u}|^2/c^2}}=\frac{c\mathbf{u}}{\sqrt{c^2+|\mathbf{u}|^2}}\\ \frac{d\mathbf{u}}{dt}=\frac{\mathbf{F}(\mathbf{x},\mathbf{u})}
{\sqrt{m^2+|\mathbf{F}(\mathbf{x},\mathbf{u})|^2/a_m^2}} =\frac{a_m\mathbf{F}(\mathbf{x},\mathbf{u})}
{\sqrt{(ma_m)^2+|\mathbf{F}(\mathbf{x},\mathbf{u})|^2}}  .
      \end{array}
    \right .
\end{equation}
If $\mathbf{F}(\mathbf{x},\mathbf{u})$ is smooth, then, by a theorem of differential equations,
a solution of the above system exists and is unique for any initial conditions $\mathbf{x}(0),\mathbf{u}(0)$.
Moreover, from (\ref{EST dym system}) it follows directly that the solution satisfies the limitations of ERD:
\[|\mathbf{v}|=\left|\frac{d\mathbf{x}}{dt}\right|\leq c,\;\;
|\mathbf{a}|=\left|\frac{d\mathbf{u}}{dt}\right|\leq a_m\,.\]
The acceleration $a$ of any massive object is limited by the maximal one, $a<a_m$. The proper velocity time trajectory of such objects is inside the "light cone" $ |\mathbf{u}|<a_m t$ in $(t,\mathbf{u})$. The zero-mass particles propagate on the boundary of the cone.

To transform the system (\ref{EST dym system}) into a Hamilton type system, we introduce the
relativistic momentum $\mathbf{p}=m\mathbf{u}$ and a phase space $(\mathbf{x},\mathbf{p})$. Then,
we can rewrite the ESR dynamics equations as a Hamilton type system on the phase space:
\begin{equation}\label{Hamilton}
   \left\{
      \begin{array}{l}
       \frac{d\mathbf{x}}{dt} =\frac{\mathbf{p}}{m}\frac{1}{\sqrt{1+|\mathbf{p}|^2/(mc)^2}}=
       \frac{c\mathbf{p}}{\sqrt{(mc)^2+|\mathbf{p}|^2}}\\ \frac{d\mathbf{p}}{dt}=\mathbf{F}(\mathbf{x},\mathbf{p})\frac{1}
{\sqrt{1+|\mathbf{F}(\mathbf{x},\mathbf{p})|^2/(ma_m)^2}}=
\frac{ma_m\mathbf{F}(\mathbf{x},\mathbf{p})}
{\sqrt{(ma_m)^2+|\mathbf{F}(\mathbf{x},\mathbf{p})|^2}}
      \end{array}
    \right .
\end{equation}
This form of the equation define also the dynamics of zero-mass particles. The equations are similar
for both $\mathbf{x}$ and $\mathbf{p}$ as suggested by Borns's reciprocity.

The system becomes a Hamilton system if there is a Hamiltonian $H(\mathbf{x},\mathbf{p})$
satisfying
\begin{equation}\label{Hamilton condition 1}
    \frac{\partial H}{\partial p_j}=\frac{p_j}{m}\frac{1}{\sqrt{1+|\mathbf{p}|^2/(mc)^2}}
\end{equation}
and
\begin{equation}\label{Hamilton condition 2}
    \frac{\partial H}{\partial r_j}=-\frac{F_j(\mathbf{x},\mathbf{p})}
{\sqrt{1+|\mathbf{F}(\mathbf{x},\mathbf{p})|^2/(ma_m)^2}}\,.
\end{equation}
The solution of (\ref{Hamilton condition 1}) is $H(\mathbf{x},\mathbf{p})=mc^2\sqrt{1+|\mathbf{p}|^2/(cm)^2}+U(\mathbf{x})$ for some function $U(\mathbf{x})$. Substituting this expression into
(\ref{Hamilton condition 2}) shows that a Hamiltonian can be found only if $\mathbf{F}(\mathbf{x},\mathbf{p})$ does not depend on $\mathbf{p}$. In this case, we have
\begin{equation}\label{potential energy}
    \frac{\partial U(\mathbf{x})}{\partial r_j}=-\frac{F_j(\mathbf{x})}
{\sqrt{1+|\mathbf{F}(\mathbf{x})|^2/(ma_m)^2}}\,,
\end{equation}
and the Hamiltonian is
\begin{equation}\label{Hamiltonian formula}
   H(\mathbf{x},\mathbf{p})=mc^2\sqrt{1+|\mathbf{p}|^2/(mc)^2}+U(\mathbf{x})\,,
\end{equation}
where $U(\mathbf{x})$ is defined by (\ref{potential energy})
The first term of the Hamiltonian coincides with the usual relativistic energy of a free particle, while the second term is the potential energy.

%

Consider first the harmonic oscillator, an object of mass $m$  affected by a force $F_j=-kx_j,\;\;j=1,2,3,$ for some constant $k$. In this case, the ESR potential energy, defined by equation (\ref{potential energy}), is
\[\frac{\partial U(\mathbf{x})}{\partial x_j}=\frac{kx_j}
{\sqrt{1+k^2|\mathbf{x}|^2/(ma_m)^2}}\,,\]
which has a solution
\begin{equation}\label{potential harmonic oscil}
  U(\mathbf{x})=\frac{(ma_m)^2}{k}( \sqrt{1+k^2|\mathbf{x}|^2/(ma_m)^2}-1),
\end{equation}
where we chose the free constant which gives $U(0)=0$.
If $k^2|\mathbf{x}|^2/(ma_m)^2\ll 1$, then expanding the square root into a power series yields
\[  U(\mathbf{x})=\frac{1}{2}k|\mathbf{x}|^2\left(1-\frac{1}{4}\frac{k^2|\mathbf{x}|^2}{(ma_m)^2} +\cdots\right),\]
which shows the approximation to non-relativistic potential energy of a harmonic oscillator.

%

Consider another example: the motion of an object of mass $m$ in a central field, with force $F_j=kx_j/|\mathbf{x}|^3$,
for some constant $k$. In this case, the ESR potential energy, defined by equation (\ref{potential energy}), is
\[\frac{\partial U(\mathbf{x})}{\partial x_j}=-\frac{kx_j}
{|\mathbf{x}|^3\sqrt{1+k^2/(|\mathbf{x}|^2ma_m)^2}}=
-k\frac{1}
{\sqrt{|\mathbf{x}|^4+k^2/(ma_m)^2}}\frac{x_j}{|\mathbf{x}|}\,,\]
which has a solution
\begin{equation}\label{potential central field}
    U(\mathbf{x})=k\left(\int_0^{\infty}\frac{ds}
    {\sqrt{s^4+k^2/(ma_m)^2}}-\int_0^{|\mathbf{x}|}\frac{ds}
    {\sqrt{s^4+k^2/(ma_m)^2}}\right)\,.
\end{equation}
For $|\mathbf{x}|\ll\sqrt{k/(ma_m)}$ we have $U(\mathbf{x})\approx -ma_m|\mathbf{x}|$, while for
$|\mathbf{x}|\gg\sqrt{k/(ma_m)}$ we have $U(\mathbf{x})\approx -k/|\mathbf{x}|+const$.

\section{The Doppler shift for a accelerated observer or source}

To define the value of the maximal acceleration, we will need a formula for the Doppler shift for a accelerated observer or source. Since Doppler shift formulas may be derived from the Lorentz transformations, similar formulas for the Doppler shift will hold for comoving accelerated systems by replacing $v/c$ with $a/a_m$.

 For this paper, we need only the longitudinal shift for radiations in the direction of the radiation. Assume that our source is radiating with frequency $\omega$ and both of direction od acceleration $a$ and the direction of radiation $\tilde{k}$ are in the $x$-direction. In the proper velocity-time representation, the radiation is described by $f(\omega t- \tilde{k} u)$. Using (\ref{coordtranspfinal2}), we get
\[ f(\omega t- \tilde{k} u)=f\left(\omega \tilde{\gamma}(t'+ \frac{au'}{a_m^2})- \tilde{k} \tilde{\gamma}(at'+u')\right)\]\[=
f\left(\tilde{\gamma}(\omega -\tilde{k}a)t'-\tilde{\gamma}(\tilde{k}-\frac{a}{a_m^2})u'\right)=f(\omega ' t'- \tilde{k}' u ').\]
From this, one gets the Doppler shift for an accelerated observer to be
\begin{equation}\label{DopplerAccel}
    \omega '=\tilde{\gamma}(\omega -\tilde{k}a)=\frac{\omega -\tilde{k}a}{\sqrt{1-\frac{a^2}{a_m^2}}}=\omega\frac{1 -\frac{|\tilde{k}|}{\omega}a}{\sqrt{1-\frac{a^2}{a_m^2}}}=\omega\frac{1 -\frac{a}{a_m}}{\sqrt{1-\frac{a^2}{a_m^2}}}.
\end{equation}
If $\frac{a}{a_m}\ll 1$, we get
\begin{equation}\label{DopplerAprox}
    \omega '=\omega \left(1-\frac{a}{a_m}\right)
\end{equation}

\section{ K\"{u}ndig's experiment and its consequences}
K\"{u}ndig's experiment (K\"{u}ndig (1963)) measured the transverse Doppler effect in a rotating disk by means of the M\"{o}ssbauer effect.
 \begin{figure}[h!]
  \centering
\scalebox{0.8}{\includegraphics{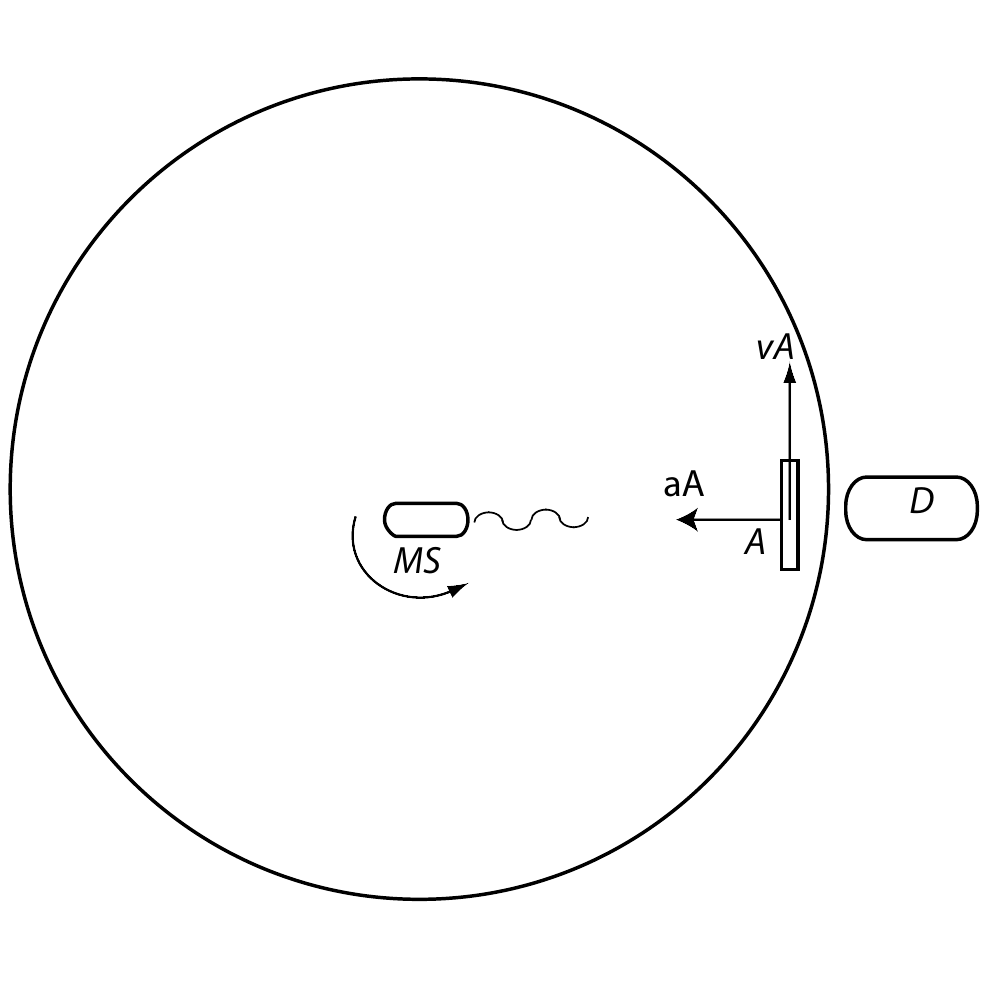}}
  \caption{The rotating disk in the K\"{u}ndig's experiment of transverse Doppler effect}
\end{figure}
In this experiment, the distance from the center of the disk to the absorber was $R=9.3 cm$, and the rotation velocity varied between $300-35000$ $rpm$. The velocity
$\mathbf{v}=R\omega $ of the absorber is perpendicular to the radius, the radiation direction. K\"{u}ndig expected to measure the transverse Doppler effect by measuring
the relative energy shift, which, by relativity, should be
\begin{equation}\label{doplerShift}
  \frac{\triangle E}{E}\approx -\frac{R^2\omega ^2}{2c^2},
\end{equation}
where $E$ is the photon energy as judged from its frequency.

 Let us introduce a constant $b$ such that
 \begin{equation}\label{b def}
   \frac{\triangle E}{E}= -b\frac{R^2\omega ^2}{2c^2}.
\end{equation}
K\"{u}ndig's analysis of experimental result led
 \begin{equation}\label{Kundigres}
  b=1.0065\pm 0.011,
\end{equation}
 which was claimed to be in full agreement with the expected time dilation.

However, Kholmetskii \textit{et al} (2008) found an error in the data processing of the results of K\"{u}ndig's experiment. They corrected the error and recalculated the results for three different rotation velocities for which the authors of the experiment provided all the necessary data. After their corrections,
the average value of $b$ is
  \begin{equation}\label{KholRes}
   b=1.192\pm 0.03,
  \end{equation}
 which does \textit{not} agree with (\ref{doplerShift}). The individual values of $b$ for each of the three velocities are given in Table 1.
\begin{table}[h!]
  \centering
 \begin{tabular}{|c|c|}
   \hline
   Speed of rotor (rpm) & b \\
    \hline
   11000& $1.178\pm 0.07$ \\
   21000 & $1.172\pm 0.02$ \\\
   31000 & $1.221\pm 0.02$ \\
   \hline
 \end{tabular}
 \caption{ The values of $b$ from 3 experiments of K\"{u}ndig recalculated by
 Kholmetskii \textit{et al}.
 }\label{cascade1end}
 \end{table}

We now use the above results and show that the Clock Hypothesis is not valid. This, in turn, leads us to predict a maximal acceleration.The clock hypothesis has been tested and was found to be valid to great accuracy. See, for example, the experiment of Bailey J. \textit{et al} (1977) for measurements of the time dilation for muons. This implies that if there is a maximal acceleration, it must be larger than the acceleration in that experiment.

The absorber is rotating. Hence, its velocity is perpendicular to the radius, and its acceleration is toward the source of radiation. Let $K$ denote the inertial frame of the lab. We can attach an accelerated system $K''$ to the absorber. Introduce an inertial frame $K'$ co-moving with the absorber. The frame $K'$ moves parallel to $K$ with constant velocity $\mathbf{v}=R\omega$. The time dilation between $K$ and $K'$ is given by the transverse Doppler effect, as in (\ref{doplerShift}). If the Clock Hypothesis, claiming that there is no effect on the rate of the clock due to acceleration, is valid, then there is no change in time from system $K'$ to $K''$. As a result, formula (\ref{doplerShift}) should
also hold for time dilation between $K$ and $K''$. However, by (\ref{KholRes}), this is not the case, with a deviation exceeding almost 20 times the measuring error.

In  K\"{u}ndig's experiment, the system $K''$ moves with acceleration $a=R\omega^2$ toward the source. Thus, time transformations between the inertial system $K'$ and the accelerated co-moving system $K''$ will be given by a longitudinal Doppler shift given by (\ref{DopplerAprox}).
Thus, time transformations between system $K$ and $K''$ are the product of the transverse Doppler transformation between
$K$ and $K'$ and the longitudinal Doppler effect due to the acceleration of $K'$ with respect to $K''$. We have
\[  \left( 1-\frac{R\omega^2}{a_m}\right) \sqrt{1-\frac{R^2\omega^2}{c^2}}\approx\left( 1-\frac{R\omega^2}{a_m}\right)\left( 1-\frac{R^2\omega^2}{2c^2}\right)\]\[\approx1-\frac{R\omega^2}{a_m} -\frac{R^2\omega^2}{2c^2}=1-\left( 1+\frac{2c^2}{Ra_m}\right)\frac{R^2\omega^2}{2c^2}.
\]
This implies that
\begin{equation}\label{bformula}
    b=1+\frac{2c^2}{Ra_m}.
\end{equation}
Notice that the calculated value of $b$ is independent of the speed of rotation. This agrees approximately with the data in Table 1.

By substituting the observed time dilation in  K\"{u}ndig's experiment from (\ref{KholRes}) and $R=0.093$, we get
\[b=1+\frac{2c^2}{Ra_m}=1.192\pm 0.03,\]
implying that
\begin{equation}\label{maxaccelcalc}
   a_m=\frac{2c^2}{R(0.192\pm 0.03)}=(112\pm 7)c^2=(1.006\pm0.063)10^{19}m/s^2.
\end{equation}

In Friedman (2009) is proposed an experiment test the clock hypothesis and to measure the maximal acceleration

\section{Discussion}

We have seen that the value of the maximal acceleration is very large. This explains why it was not detected in other experiments which did verify time dilation.
K\"{u}ndig's experiment was designed to measure a transversal  Doppler effect of order $v^2/c^2$ due to the velocity of the absorber. On the other hand, the acceleration was in the direction of the radiation, causing a longtitudal Doppler shift of order $a/a_m$. Thus, we were able to observe the influence of the acceleration even though the maximal acceleration has a value of order $c^2/R$. The maximal accelerations calculated for each of the three experiments of K\"{u}ndig are
 close to the value of the maximal acceleration predicted by the average value.

The existence of a maximal acceleration was conjectured by Caianiello (1981), based on time-energy uncertainty in quantum mechanics. It has also been conjectured by Schuller (2002), using Born-Infeld theory. Schuller (2002) obtained transformation formulas between accelerated systems similar to ours. But his maximal acceleration is a proper maximal accelerations $d^2x/d\tau^2$, which is larger them our acceleration, and differs from particle to particle. Based on an experiment of  Bailey J. \textit{et al} (1977), Schuller estimates his maximal acceleration to be larger then $a>2.5\cdot10^{18}g$, but for our maximal acceleration this experiment give an estimate $a_m>1.9\cdot10^{17}g$.
 Since this estimate is close to the maximal acceleration, improved muon lifetime experiments could provide another way to determine the maximal acceleration.
For a long time, B. Mashhoon argued against the clock hypothesis and developed nonlocal transformations for accelerated observers (see the review article Mashhoon (2008)). Our approach treats the problem differently.

Note that the existence of a maximal acceleration will necessitate the modification of the Lorentz transformations and Maxwell's equations, which allow unlimited acceleration. The major modification will come on the quantum scale. Note that at quantum system distances, the classical electromagnetic force would generate accelerations above the maximal one. For example, the classical electric field of a proton would generate a maximal acceleration on an electron positioned at a distance of about $50\dot{A}$ from it. At quantum distances, therefore, the potential of the electron will differ significantly from the classical one, and, by assuming continuity at the origin, it follows that this potential has a potential well. This may provide another model for the hydrogen atom. For an additional example, take the classical electric field of a proton, which would generate a maximal acceleration on another proton positioned at a distance of about $1\dot{A}$ from it. By assuming continuity at the origin, we obtain a potential barrier for the proton similar to the strong force at small distances.

In light of these observations, we propose the following problem:

\noindent \textbf{Problem} \textit{Is Quantum Mechanics equipped to handle the maximal acceleration?  On the other hand, can the extended relativity, preserving the limitation of the maximal acceleration, also incorporate quantum phenomena? }

\noindent As shown by Caianiello  (1981) and others, one of the main features of quantum mechanics, the uncertainty, is connected with existence of a maximal acceleration.

 The Lorentz transformations may be modified implementing methods of bounded symmetric domains, see Friedman (2004) and  Friedman and Semon (2005), for the symmetric domain of relativistically admissible velocities and accelerations. This may be done by use of the relativistic phase space introduced in Friedman (2008).

K\"{u}ndig's experiment was not designed to test our observation, and in the recalculation by
 Kholmetskii \textit{et al}, the measuring error is underestimated, because the data processing method applied is not direct. Thus, our finding is only an indication of the existence of a maximal acceleration and an estimate of its value. On the other hand, we have shown that an experiment determining the value of maximal acceleration could be done with currently available technology.
\medskip

I would like to thank A.L. Kholmetskii for directing my attention to his paper and for helpful discussions. I would also like to thank J. Bekenstein and U, Sandler for constructive remarks, Yu. Gofman for inspiring discussions and T. Scarr for editorial comments.

  \end{document}